\documentclass[sigconf,natbib=false]{acmart}
\usepackage{multirow}
\AtBeginDocument{%
  }
    
\usepackage[ruled, vlined, linesnumbered]{algorithm2e}
\allowdisplaybreaks
\copyrightyear{2022} 
\acmYear{2022} 
\setcopyright{acmlicensed}\acmConference[ICAIF '22]{3rd ACM International Conference on AI in Finance}{November 2--4, 2022}{New York, NY, USA}
\acmBooktitle{3rd ACM International Conference on AI in Finance (ICAIF '22), November 2--4, 2022, New York, NY, USA}
\acmPrice{15.00}
\acmDOI{10.1145/3533271.3561673}
\acmISBN{978-1-4503-9376-8/22/10}

\RequirePackage[
  datamodel=acmdatamodel,
  style=acmnumeric,
  ]{biblatex}

\begin{document}

\title{Equitable Marketplace Mechanism Design}

\author{Kshama Dwarakanath}
\affiliation{%
  \institution{J.P.Morgan AI Research}
  \city{Palo Alto}
  \state{California}
  \country{USA}
}
\email{kshama.dwarakanath@jpmorgan.com}

\author{Svitlana S Vyetrenko}
\affiliation{%
  \institution{J.P.Morgan AI Research}
  \city{Palo Alto}
  \state{California}
  \country{USA}
}

\author{Tucker Balch}
\affiliation{%
  \institution{J.P.Morgan AI Research}
  \city{New York}
  \state{New York}
  \country{USA}
}

\renewcommand{\shortauthors}{Dwarakanath et al.}

\begin{abstract}
We consider a trading marketplace that is populated by traders with diverse trading strategies and objectives. The marketplace allows the suppliers to list their goods and facilitates matching between buyers and sellers. In return, such a marketplace typically charges fees for facilitating trade. The goal of this work is to design a dynamic fee schedule for the marketplace that is equitable and profitable to all traders while being profitable to the marketplace at the same time (from charging fees). Since the traders adapt their strategies to the fee schedule, we present a reinforcement learning framework for simultaneously learning a marketplace fee schedule and trading strategies that adapt to this fee schedule using a weighted optimization objective of profits and equitability. We illustrate the use of the proposed approach in detail on a simulated stock exchange with different types of investors, specifically market makers and consumer investors. As we vary the equitability weights across different investor classes, we see that the learnt exchange fee schedule starts favoring the class of investors with the highest weight. We further discuss the observed insights from the simulated stock exchange in light of the general framework of equitable marketplace mechanism design.
\end{abstract}


\maketitle




\section{Introduction}\label{sec:marketplace}
We consider a marketplace with trading agents that play the roles of buyers and/or sellers of goods or services. The trading agents are of different types based on the frequency at which they trade, the size of their trading orders, whether they consistently buy or sell or both, and the strategy that they use for trading. The seller agents can range from manufacturers that offer high good volumes for sale, to individual and retail sellers that sell much less frequently. The buyer agents use the platform to compare prices available from different seller agents, and to buy goods from them. The \emph{marketplace} agent is one that facilitates trading between buyer and seller agents by providing them access to marketplace communication and an order matching engine. It also charges trading agents in the form of fees for the facilities it provides. These fees serve as profits to the marketplace agent. Further, the profits made by trading agents through goods exchange are offset by the fees paid to the marketplace. 

\emph{Wholesale} agents are both buyers and sellers of the goods that are traded in the marketplace. They are characterised by their large trading volumes and high trading frequencies. Wholesale agents can improve liquidity in the marketplace by frequent buying and selling of large volumes of goods. It is for this reason that the marketplace agent often offers fee rebates to wholesale agents for their function of ensuring the presence of orders on the opposite side of every arriving buy or sell order. There are \emph{other} agent groups in the marketplace ecosystem that trade goods based on their perception of the long term value of a good, or market momentum signals. \emph{Consumer} agents are characterized by trading on demand. They arrive to the marketplace at random times and trade smaller volumes without using informed trading strategies. They can therefore potentially trade at inferior prices, hence, raising questions about marketplace equitability to consumer agents. Consider an example where a consumer agent needs to trade during a period of marketplace distress when there is little supply of goods offered by wholesale agents. Under such conditions, the consumer agent might trade at an inferior price, resulting in an execution that may be perceived as inequitable as compared to those of other agents who were luckier to trade in a more stable marketplace.

Examples of marketplace ecosystems that can be described by the above framework include e-commerce trading platforms as well as stock exchange markets. For instance, wholesale agents such as publishing houses as well as other small seller agents (such as bookstores) can list books for a fee on e-commerce trading marketplaces, which allows them to procure books to individual consumer agents and other buyers. In stock exchange markets, market makers provide liquidity to the exchange on both buy and sell sides of the market for exchange incentives. This action enables market makers to profit, and other market agents to trade assets with each other. Consumer agents such as individuals who trade on demand without sophisticated trading strategies and technology can be vulnerable to rapid price changes and volatility. 

Simulations have previously been used to answer questions that are of interest to participants of financial markets, hedge funds, banks and stock exchanges. In \cite{nasdaq_tick_size}, the authors investigated the use of an intelligent tick structure that modified the currently constant tick size for all stocks to having different tick sizes for different stocks. Fairness and equitability in markets have become increasingly important as described in \cite{sec_options,sec_securities}. In this paper, we investigate the impact of a reduction in marketplace fees charged to wholesale agents on equitability outcomes to consumer agents in a marketplace simulator. We show that such fee reductions incentivise wholesale agents to enable equitable provision of goods to consumer agents in the marketplace (see Figure \ref{fig:marketplace}). Specifically, we demonstrate that an equitable marketplace mechanism can be enabled by a dynamic marketplace fee policy derived by reinforcement learning in a simulated stock exchange market.

\section{Background and related work}

\subsection{Equitability Metric}

Equitability has conventionally been studied in political philosophy and ethics \cite{johnrawls}, economics \cite{hmoulin}, and public resource distribution \cite{peytonyoung}. In recent times, there has been a renewed interest in quantifying equitability in classification tasks \cite{dwork2012fairness}. Literature on fairness in machine learning studies two main notions of equitability: group fairness and individual fairness. Group fairness ensures some form of statistical parity (e.g. between positive outcomes, or errors) for members of different groups. On the other hand, individual fairness ensures that individuals who are `similar' with respect to the task at hand receive similar outcomes \cite{binns2020apparentfairness}. In \cite{dwarakanath2021profit}, the authors studied the effect a market marker can have on individual fairness for consumer traders by adjusting its parameters. A negative correlation was observed between the profits of the market maker and equitability. Hence, the market maker incurs a cost while enabling individual fairness to consumer traders. This stirs up the idea of designing a marketplace in which the market maker can be compensated by the exchange for making equitable markets.

\begin{figure}[tb]
    \centering
    \includegraphics[width=\linewidth]{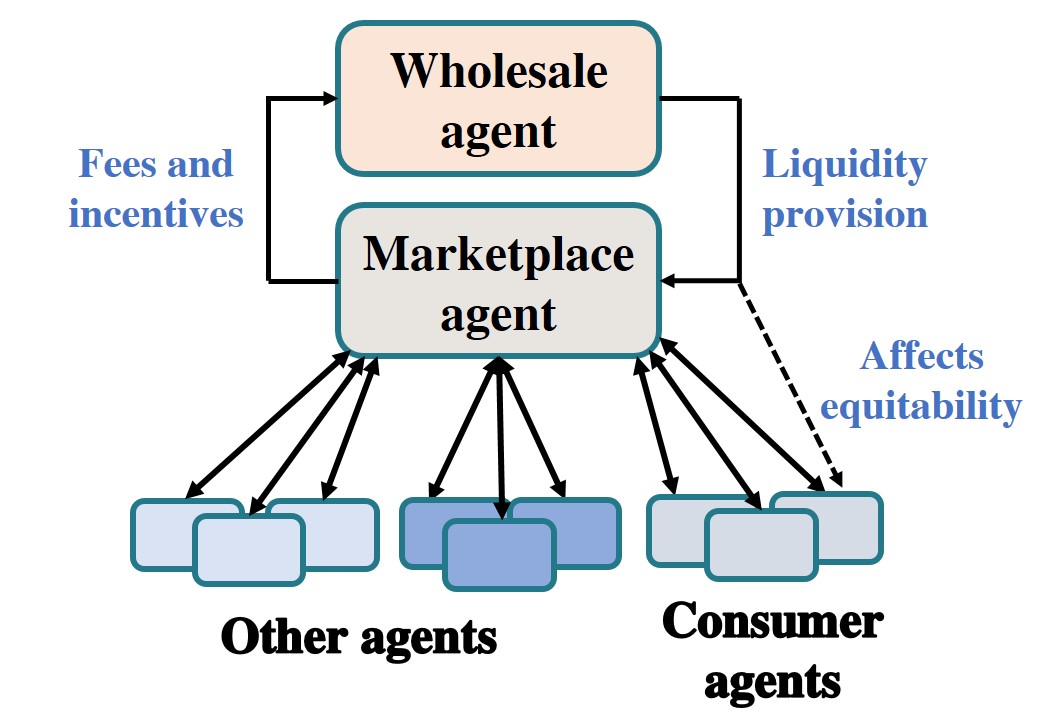}
    \caption{Schematic of marketplace ecosystem}
    \label{fig:marketplace}
\end{figure}

In this paper, we are interested in understanding the effects of marketplace fees on equitability to trading agents. We draw from the entropy metric used in \cite{dwarakanath2021profit} to measure individual fairness within a single group of (consumer) traders. We seek an equitability metric that can capture equitability both within each group as well as that across groups. The authors of \cite{cowell1981gei,algo_unfairness} describe the family of generalized entropy indices (GEI) that satisfy the property of subgroup-decomposability, i.e. the inequity measure over an entire population can be decomposed into the sum of a between-group inequity component (similar to group fairness metrics in \cite{dwork2012fairness,aif360}) and a within-group inequity component (similar to the individual fairness entropy metric used in \cite{dwarakanath2021profit}). Given observations $Y=\lbrace y_1,y_2,\cdots,y_n\rbrace$ of outcomes of $n$ agents, the generalized entropy index is defined for parameter $\kappa\neq0,1$ as
\begin{align}
    \textnormal{GE}_{\kappa}(Y)=\frac{1}{n\kappa(\kappa-1)}\sum_{i=1}^n\left(\left(\frac{y_i}{\mu}\right)^\kappa-1\right)\label{eq:gei_defn}
\end{align}
where $\mu:=\frac{1}{n}\sum_{i=1}^{n}y_i$ is the average outcome. Note that $\textnormal{GE}_{\kappa}(Y)$ is a measure of inequity with the most equitable scenario resulting from $y_i=c$ for all $i=1,2,\cdots,n$. If we think of $y_i$ to denote the profit of trading agent $i$, then the most equitable scenario corresponds to having equal profits for all agents. 

With the population divided into $G$ groups of agents, one can decompose the right hand side of equation (\ref{eq:gei_defn}) as \begin{align}
   \textnormal{GE}_\kappa(Y)&=\sum_{g=1}^{G}\frac{n_g}{n}\left(\frac{\mu_g}{\mu}\right)^\kappa \textnormal{GE}_\kappa(Y_g)\label{eq:gei1}\\
   &+\sum_{g=1}^{G}\frac{n_g}{n\kappa(\kappa-1)}\left(\left(\frac{\mu_g}{\mu}\right)^\kappa-1\right)\label{eq:gei2}
\end{align}
where $Y_g=\lbrace y_i:i\in g\rbrace$ is the set of outcomes of agents in group $g$, $n_g$ is the number of agents in group $g$, and $\mu_g:=\frac{1}{n_g}\sum_{i\in g}y_i$ is the average outcome in group $g$. Then, the term on the right of (\ref{eq:gei1}) captures the within-group inequity (similar to the entropy metric for individual fairness in \cite{dwarakanath2021profit}) for group $g$ weighted by its population. And, the term in (\ref{eq:gei2}) captures the between-group inequity by comparing the average outcome in the entire population against that in group $g$. 

We propose a weighted version of (\ref{eq:gei1})-(\ref{eq:gei2}) with weight $w=\lbrace w_g:g=1,2,\cdots,G\rbrace$ where $w_g\geq0$ for all $g=1,2,\cdots,G$ and $\sum_{g=1}^{G}w_g=1$ as
\begin{equation}
    \begin{aligned}
    \textnormal{GE}^w_\kappa(Y)&=\sum_{g=1}^Gw_g\cdot\frac{n_g}{n}\left(\frac{\mu_g}{\mu}\right)^\kappa \textnormal{GE}_\kappa(Y_g)\\
   &+\sum_{g=1}^Gw_g\cdot\frac{n_g}{n\kappa(\kappa-1)}\left(\left(\frac{\mu_g}{\mu}\right)^\kappa-1\right) 
    \end{aligned}\label{eq:w_gei}
\end{equation}
Note that the equitability metric defined in (\ref{eq:w_gei}) provides extended flexibility to the original definition (\ref{eq:gei_defn}) by enabling the user to focus on a specific agent group $l$ by setting $w_l=1$ and $w_g=0$ for all $g\neq l$. For ease of notation, we establish the following group correspondence for the three types of trading agents in our marketplace described in section \ref{sec:marketplace}. Let $g=1$ correspond to wholesale agents, $g=2$ to consumer agents and $g=3$ to other agents. We use the negative of (\ref{eq:w_gei}) as the metric for equitability going ahead.

\subsection{Reinforcement Learning}
Our marketplace ecosystem consists of multiple interacting trading agents as in Figure \ref{fig:marketplace}. 
Such an ecosystem is well modeled as a multi-agent system -- a system comprised of multiple autonomous agents interacting with each other in a common environment which they each observe and act upon. The behaviours of these agents can be defined beforehand using certain rules or expert knowledge, or learnt on the go. Reinforcement learning (RL) has become a popular approach to learn agent behavior given certain objectives that are to be improved upon \cite{sutton2018reinforcement,kaelbling1996reinforcement}. An RL agent seeks to modify its behaviour based on rewards received upon its interaction with its dynamic environment. 
There exist well-understood algorithms with their convergence and consistency properties well studied for the single-agent RL task. 

An environment with multiple learning agents is modeled in the form of Markov Games (MGs) or stochastic games \cite{zhang2021marl,shapley1953stochastic}. An MG is a tuple $(\mathcal{N},\mathcal{S},\lbrace\mathcal{A}_i:i=1,2,\cdots,n\rbrace,\mathcal{T},\lbrace R_i:i=1,2,\cdots,n\rbrace,\gamma,T)$ comprising the set $\mathcal{N}=\lbrace1,2,\cdots,n\rbrace$ of all agents, the joint state space, action spaces of all agents, a model of the environment giving the probability of transitioning from one joint state to another given the actions of all agents, the reward functions of all agents, discount factor and the time horizon respectively \cite{zhang2021marl}. The goal of each agent is to maximize the expected sum of its own discounted rewards that now depend on the actions of other agents as well. While it is tempting to use RL for multi-agent systems, it comes with a set of challenges. The main challenge being the presence of multiple agents that are learning to act in presence of one another \cite{busoniu2008comprehensivemarl}. We deal with this by adopting an iterative learning framework where our learning agents taking turns to update their value functions while the other learning agents keep their policies fixed. Such an iterative learning framework was used in \cite{zheng2020aieconomist} to simultaneously learn economic actors and a tax planner in an economic system.

The general framework of using RL for mechanism design was previously considered in \cite{rl_mechanismdesign}. Mechanism design using RL for e-commerce applications was studied in \cite{mechanism_ecommerce}. In this paper, we show how to use RL for equitable marketplace mechanism design with our discussion focused on financial markets.

\subsection{Stock exchange markets}
We now concentrate on stock exchange markets (such as Nasdaq or New York Stock Exchange) which can be viewed as instances of our generic trading marketplace with stocks being the goods traded between agents. 
Stock trading agents can belong to many categories: market makers, consumer investors, fundamental investors, momentum investors, etc. Market makers are investors that are obliged to continuously provide liquidity to both buyers and sellers regardless of market conditions. 
They act as both buyers and sellers of the stock and have more frequent trades with larger order volumes as compared to the other categories of investors. 

Fundamental investors and momentum investors use the exogenous stock value or long term averages of the stock to trade unilaterally (buy or sell, not both) at different times in the trading day; they also have more frequent trades (albeit unilateral) than the category of consumer investors, who trade purely based on demand without any other considerations \cite{kyle1985continuous}. 
Irrespective of type, the objective of all market investors is to make profits from their trading actions. 

The aforementioned investor categories can be mapped to our marketplace ecosystem as follows: exchange (marketplace agent), market makers (wholesale agents), consumer investors (consumer agents) and value and momentum investors (other agents) - see Figure~\ref{fig:marketplace} for agent categories. 
The exchange charges investors fees for the facilities it provides on its platform. These fees typically differ based on investor category, and serve as profits for the exchange. Direct (regular) stock exchanges such as NYSE, Nasdaq and Cboe provide incentives to market makers for liquidity provision \cite{lightspeed_inverted}. On the contrary, inverted stock exchanges such as NYSE National, Nasdaq BX, Cboe EDGA and Cboe BYX charge market makers for providing liquidity. The reasons for such fee structures range from ensuring market efficiency to faster order fills in different exchanges \cite{nasdaq_inverted,cboe_inverted}.

\begin{figure}[tb]
    \centering
    \includegraphics[width=\linewidth]{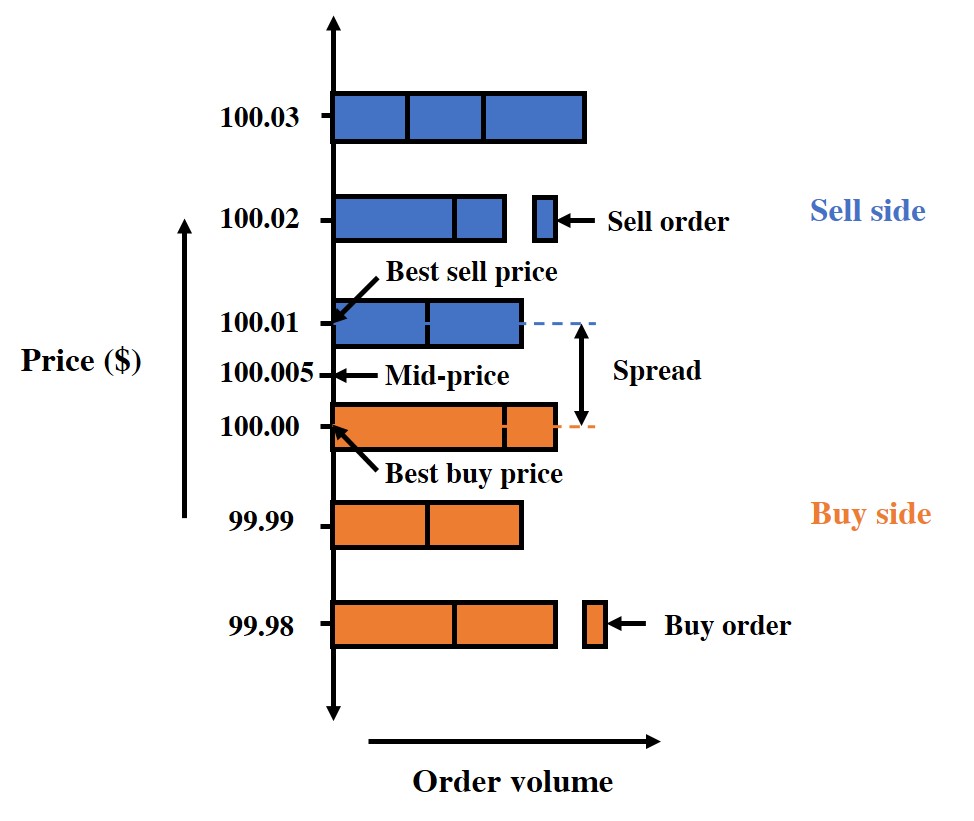}
    \caption{Snapshot of order queue maintained by the exchange}
    \label{fig:lob}
\end{figure}

\subsection{Simulator}\label{subsec:simulator}
In order to play out the interactions between agents in a stock exchange market, we employ a multi-agent exchange market simulator called ABIDES \cite{abides,amrouni2021abides}. ABIDES provides a selection of background trading agents with different trading behaviors and incentives. The simulation engine manages the flow of time and handles all inter-agent communication. The first category of simulated trading agents is that of {\it{market makers}} (denoted MMs henceforth) that continuously quote prices on both the buy and sell sides of the market, and earn the difference between the best buy and sell prices if orders execute on both sides (see Figure \ref{fig:lob}). MMs act as intermediaries and essentially eliminate `air pockets' between existing buyers and sellers. In this work, we define a MM by the stylized parameters that follow from its regulatory definition \cite{wah2017welfare,chakraborty2011market}. At every time $t$, the MM places new price quotes of constant order size $I$ at $d$ price increments around the current stock price $p_t$ in cents i.e., it places buy orders at prices $p_t-h-d, \ldots, p_t-h$ and sell orders at prices $p_t + h, \ldots, p_t + h +d$, where $d$ is the depth of placement and $h$ is the half-spread chosen by the MM at time $t$. Figure \ref{fig:lob} shows an example snapshot of orders collected at the exchange with $p_t=10000.5$ cents, $h=0.5$ cents displaying $d=2$ levels on either side of the current stock price. Each blue/orange rectangle represents a block of sell/buy orders placed in the order queue.

ABIDES also contains other strategy-based investors such as fundamental investors and momentum investors. The fundamental investors trade in line with their belief of the exogenous stock value (which we call fundamental price), without any view of the market microstructure \cite{kyle1985continuous}. 
In this paper, we model the fundamental price of an asset by its historical price series. Each fundamental investor arrives to the market according to a Poisson process, and chooses to buy or sell a stock depending on whether it is cheap or expensive relative to its noisy observation of the fundamental. On the other hand, the momentum investors follow a simple momentum strategy of comparing a long-term average of the price with a short-term average. If the short-term average is higher than the long-term average, the investor buys since the price is seen to be rising. And, vice-versa for selling. 

Further, ABIDES is equipped with {\it{consumer investors}} that are designed to emulate consumer agents who trade on demand. Each consumer investor trades once a day by placing an order of a random size in a random direction (buy or sell). 

\section{Problem Setup}\label{sec:problem_setup}

In this paper, we take a mechanism design based approach that uses RL to derive a dynamic fee schedule optimizing for equitability of investors as well as for profits of the exchange. The dynamic nature of the fee schedule is inspired by the idea that exchange fees at any time during a trading day must be contingent on the current market conditions. An important point regarding the use of dynamic exchange fee schedules is that other investors could change their trading strategies in response to varying fees. Therefore, we consider an RL setup with two learning agents interacting with each other. We use ABIDES as a stock exchange simulation platform with an exchange agent that learns to update its fee schedule, and a MM agent that learns to update its trading strategy (see Figure \ref{fig:marketplace} for schematic). The remaining investors have static rule-based policies that do not use learning. We formulate this learning scenario as an RL problem by representing the exchange with investors as a Markov Game (MG). 

\subsection{State of Markov Game}

The state for our MG captures the {\bf{shared states}} of the learning MM, learning exchange and the market comprising other (non-learning) investors as \begin{align}
    s=\begin{bmatrix}inventory & fee & incentive & market\ signals\end{bmatrix}\label{eq:state}
\end{align}
where $inventory$ is the number of shares of stock held in the MM's inventory, $fee$ refers to exchange trading fees per unit of stock charged from liquidity consumers such as consumer investors, $incentive$ refers to exchange incentives per unit of stock given out to liquidity providers for their services. By convention, negative values for $fee$ and $incentive$ imply providing rebates to liquidity consumers, and charging fees from liquidity providers respectively. $market\ signals$ contains signals such as 
\begin{align}
imbalance=\frac{total\ buy\ volume}{total\ buy\ volume +total\ sell\ volume}\nonumber
\end{align}
which is the volume imbalance in buy and sell orders in the market, $spread$ which is the difference between best sell and best buy prices, and $midprice$ which is the current mid-point of the best buy and sell prices of the stock (also called the stock price). Although the exchange may have access to the $inventory$ state for the MM, we design its policy to only depend on the latter three states. 

\subsection{Actions and rewards of the learning MM}

The \textbf{actions} of the learning MM are comprised of the stylized parameters of $half-spread$ and $depth$ that define the price levels at which the MM places buy and sell orders (as described in section \ref{subsec:simulator}), and are denoted by\begin{align}
 a_{\textnormal{MM}}=\begin{bmatrix}half-spread&depth\end{bmatrix}\nonumber
\end{align} 

While the MM profits from its trading actions, it also receives incentives from the exchange for all units of liquidity provided (negative values for which correspond to paying out fees to the exchange). Therefore, we define the \textbf{reward} $R_{\textnormal{MM}}$, that captures all MM profits and losses, by\begin{equation}
    \begin{aligned}
    R_{\textnormal{MM}}&=\textnormal{Trading profits}\\
    &+\lambda\cdot \left(incentive\times\textnormal{units of liquidity provided by MM}\right)
    \end{aligned}\label{eq:r_mm}
\end{equation}
where $\lambda\geq0$ is a weighting parameter for the importance given by the MM to exchange incentives. Note that although it makes monetary sense to have $\lambda=1$, one can theoretically examine the effects of varying $\lambda$ across other values, since the reward function in RL does not need to exactly map to monetary profits. The objective of a reinforcement learning MM is to find a policy by maximizing the expected sum of discounted rewards (\ref{eq:r_mm}).

\subsection{Actions and rewards of the learning exchange}

The \textbf{actions} for the exchange involve specifying fees and incentives per unit of stock placed by liquidity consumers and providers respectively denoted by \begin{align}
    a_{\textnormal{Ex}}=\begin{bmatrix}fee & incentive \end{bmatrix}\nonumber
\end{align} to entirely specify the next states of $fee$ and $incentive$ in (\ref{eq:state}).

In order to write down the {\bf{rewards}} for the exchange, we need a way to numerically quantify equitability alongside its profits. We use the negative of the weighted generalized entropy index defined in (\ref{eq:w_gei}), with the outcome $y_i$ for each investor $i\in\lbrace1,2,\cdots,n\rbrace$ being its profits at the end of the trading day. Since investors can also make loses, $y_i$ and hence $\mu$ and $\mu_g$ can take on negative values in (\ref{eq:w_gei}). This restricts the choice of $\kappa$ to even values. We choose $\kappa=2$ as in \cite{algo_unfairness} since higher values give spiky values for $\textnormal{GE}^w_\kappa$ hindering learning . For this work, we are interested in weights of the form $w=\begin{bmatrix}\beta&1-\beta&0\end{bmatrix}$ that look at equitability only to MMs and consumer investors for ease of understanding, with $\beta$ called the GEI weight.

Although the trading agents arrive at random times during a trading day, the equitability reward (\ref{eq:w_gei}) computed at the end of a trading day can be distributed throughout the day as follows. Define the equitability reward computed at every time step $t\in[T]$ to be the change in (\ref{eq:w_gei}) from $t-1$ to $t$ as
\begin{align}
R_\mathrm{Equitability}^t=
-\textnormal{GE}^\beta_2\left(Y^{t}\right)+\textnormal{GE}^\beta_2\left(Y^{t-1}\right)\label{eq:r_ex_equitability}
\end{align}
where $Y^t$ is the vector of profits for all investors $\lbrace1,2,\cdots,n\rbrace$ up to time $t$ and $\textnormal{GE}^\beta_2\left(Y^{0}\right):=0$. 

The profits made by the exchange are given by the difference between the fees received from liquidity consumers and the incentives given out to liquidity providers over all traded units of stock as
\begin{align}
    R_{\textnormal{Profits for Ex}}&=\textnormal{Fees}-\textnormal{Incentives}\label{eq:r_ex_profits}\\
    &=fee\times\textnormal{units of liquidity consumed}\nonumber\\
    &-incentive\times\textnormal{units of liquidity provided}\nonumber
\end{align}

Having quantified the two rewards of profits and equitability for the exchange, we use a weighted combination of the two as the {\bf{rewards}} for our learning exchange\begin{align}
    R_{\textnormal{Ex}}=R_{\textnormal{Profits for Ex}}+\eta\cdot R_\mathrm{Equitability}\label{eq:r_ex}
\end{align}
where $\eta\geq0$ is a parameter called the equitability-weight, that has the interpretation of monetary benefits in \$ per unit of equitability. (\ref{eq:r_ex}) is also motivated from a constrained optimization perspective as being the objective in the unconstrained relaxation \cite{boyd2004convex} of the problem $\max R_{\textnormal{Profits for Ex}}\textnormal{ s.t. }R_\mathrm{Equitability}\geq c$. With the rewards defined in equations (\ref{eq:r_ex_equitability})-(\ref{eq:r_ex}) and discount factor $\gamma=1$, the RL objective for the exchange can be written as
\begin{align}
    \mathbb{E}\left[\sum_{t=1}^T\left(\textnormal{Fees}-\textnormal{Incentives}\right)\right]-\eta\cdot\mathbb{E}\left[\textnormal{GE}^\beta_2\left(Y^{T}\right)\right]\nonumber
\end{align}
where $\mathbb{E}[X]$ denotes the expected value of a random variable $X$. Hence, the objective of the equitable exchange is to learn a fee schedule that maximizes its profits over a given time horizon, while minimizing inequity to investors. 


Having outlined our MG, we estimate the optimal policy using both tabular Q Learning (QL) \cite{watkins1992q} as well as the policy gradient method called Proximal Policy Optimization (PPO) from the library RLlib \cite{rllib}. Tabular QL estimates the optimal Q functions for the exchange and MM that are subsequently used to compute policies determining the dynamic fee schedule and MM trading strategy respectively.

\section{Experiments}
Given the MG formulated in the previous section, we try using tabular QL with discretized states as well as the policy gradient method called Proximal Policy Optimization (PPO) from the RLlib package \cite{ppo,rllib} with continuous states to estimate policies for the learning MM and Exchange (denoted Ex/EX). 

\subsection{Numerics}
The time horizon of interest is a single trading day from 9:30am until 4:00pm. Therefore, we set $\gamma=0.9999$ to ensure that traders do not undervalue money at the end of the trading day as compared to that at the beginning. Both the learning MM and Exchange take an action every minute giving $T=390$ steps per episode. We also normalize the states and rewards to lie within the range $[-1,1]$. The precise numerics of our learning experiments that are common to both tabular QL and PPO algorithms are given in Table \ref{tab:expt_numerics}. The values for every $(fee,incentive)$ pair are to be read as follows\footnote{The fees charged and incentives given out by real exchanges are of the order of $0.10$ cents per share \cite{nyse_prices} informing our choice of exchange actions listed in Table \ref{tab:expt_numerics}.}. If $(fee,incentive)=(0.30,0.25)$ cents, the exchange would charge 0.30 cents per trade executed by a liquidity consumer, and provide 0.25 cents of incentives per trade executed by a liquidity provider. Similarly, if $(fee,incentive)=(-0.30,-0.25)$ cents, the exchange would provide 0.30 cents of rebate per trade executed by a liquidity consumer, and charge 0.25 cents per trade executed by a liquidity provider.

\subsection{Training and convergence}

To make our problem suited to the use of tabular QL, we discretize our states by binning them. An important requirement for the functioning of the tabular QL algorithm is that there be enough visitations of each (state,action) pair. Accordingly, we pick our state discretization bins by observing the range of values taken in a sample experiment. We use an $\epsilon$ - greedy approach to balance exploration and exploitation. Additionally, since convergence of tabular QL relies on having \emph{adequate} visitation of each (state, action) pair, training training is divided into three phases - pure exploration, pure exploitation and convergence phases. During the pure exploration phase, $\alpha_n$ and $\epsilon_n$ are both held constant at high values to facilitate the visitation of as many state-action discretization bins as possible. During the pure exploitation phase, $\epsilon_n$ is decayed to an intermediate value while $\alpha_n$ is held constant at its exploration value so that the Q Table is updated to reflect the one step optimal actions. After the pure exploration and pure exploitation phases, we have the learning phase where both $\alpha_n$ and $\epsilon_n$ are decayed to facilitate convergence of the QL algorithm. The precise numerics specific to our tabular QL experiments are given in Table \ref{tab:tab_expt_numerics}. 

\begin{table}[tb]
    \allowdisplaybreaks
    \centering
    \begin{tabular}{|p{0.4\linewidth}|p{0.55\linewidth}|}\hline
        Total \# of training episodes & 2000\\\hline
        $half-spread$ & $\lbrace0.5,1.0,1.5,2.0,2.5\rbrace$cents\\\hline
        $depth$ & $\lbrace1,2,3\rbrace$cents\\\hline
        \multirow{3}{0.4\linewidth}{$(fee,incentive)$} &
        $\lbrace(0.30,0.30),(0.30,0.25),$\\
        &$(0.25,0.30),(-0.30,-0.30),$\\
        &$(-0.30,-0.25),(-0.25,-0.30)\rbrace$cents\\\hline
        $\gamma$ & $0.9999$\\\hline
        $T$ & $390$\\\hline
        $G$ & $3$\\\hline
        $\kappa$ & $2$\\\hline
        $\lambda$ & $1.0$\\\hline
        $\beta$ & $\lbrace0.0,0.3,0.5,0.6,1.0\rbrace$\\\hline
        $\eta$ & $\lbrace0,1,10,100,1000,10000\rbrace$\\\hline
    \end{tabular}
    \caption{Numerics common to tabular QL and PPO experiments}
    \label{tab:expt_numerics}
\end{table}

\begin{table}[tb]
    \centering
    \begin{tabular}{|p{0.5\linewidth}|p{0.45\linewidth}|}\hline
        \# of pure exploration episodes & 800 \\\hline
        \# of pure exploitation episodes & 400 \\\hline
        \# of convergence episodes & 800 \\\hline
        $\alpha_0=\cdots=\alpha_{399}=\cdots=\alpha_{599}$  & $0.9$\\\hline
        $\alpha_{999}$ & $10^{-5}$\\\hline
        $\epsilon_0=\epsilon_1=\cdots=\epsilon_{399}$ & $0.9$\\\hline
        $\epsilon_{599}$ & $0.1$ \\\hline
        $\epsilon_{999}$ & $10^{-5}$\\\hline
        \end{tabular}
    \caption{Numerics specific to tabular QL experiments}
    \label{tab:tab_expt_numerics}
\end{table}

\begin{figure*}[tb]
    \centering
    \includegraphics[width=\linewidth]{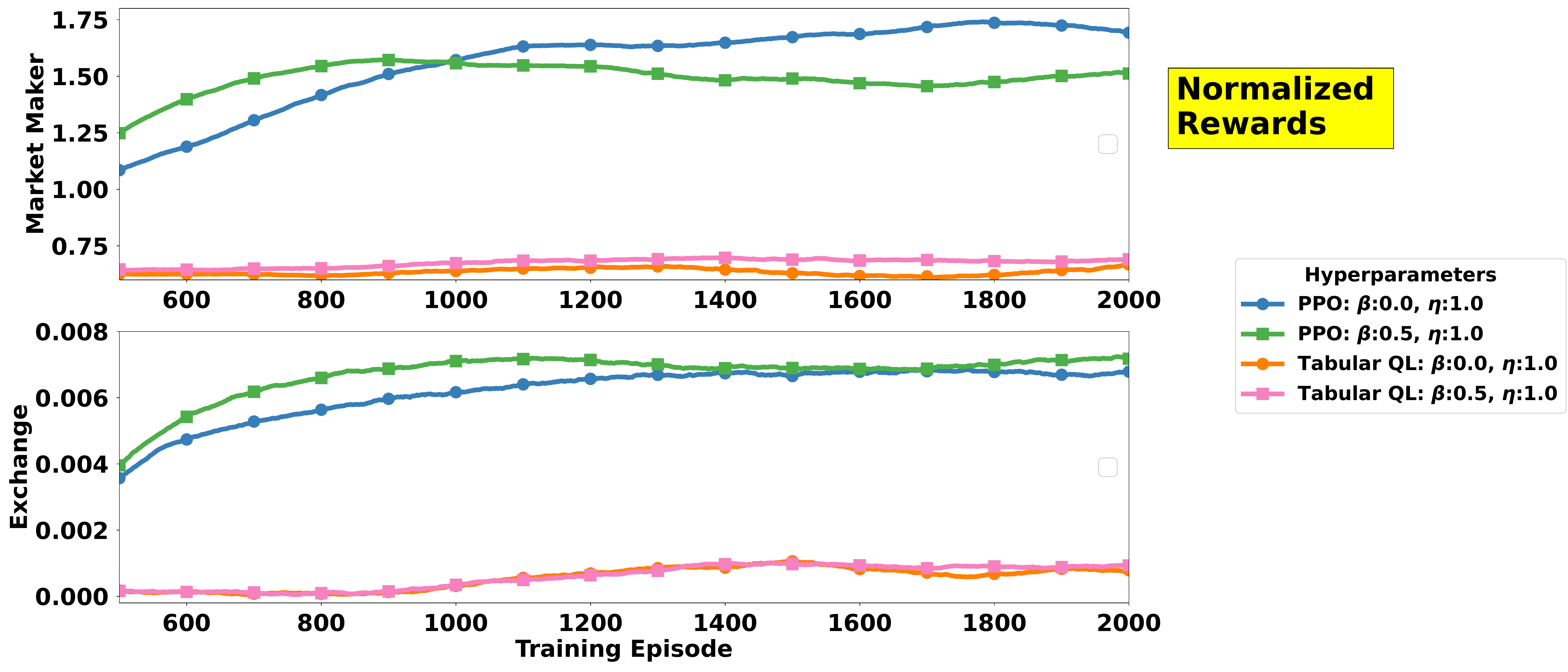}
    \caption{Comparison of training rewards for tabular QL and PPO}
    \label{fig:deep_v_tab}
\end{figure*}

\begin{figure*}[tb]
    \centering
    \includegraphics[width=\linewidth]{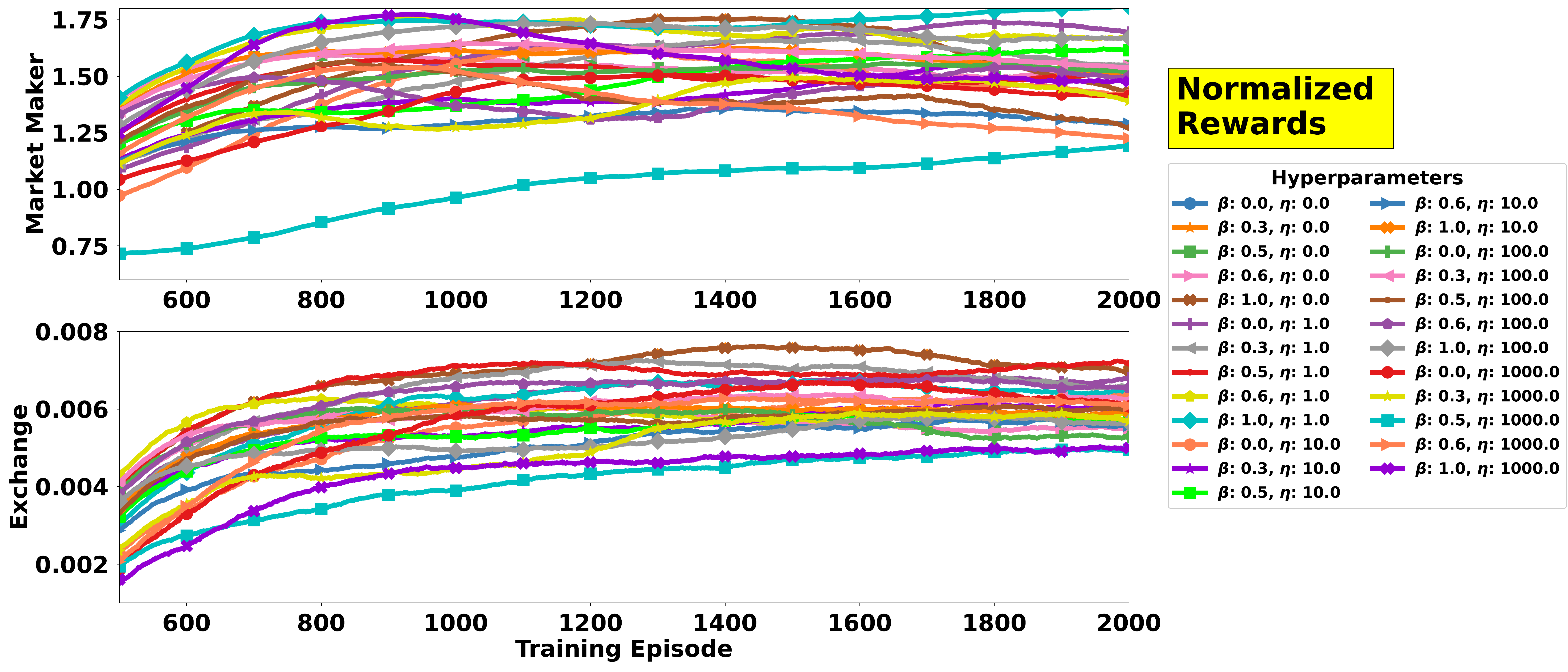}
    \caption{Training rewards with PPO}
    \label{fig:training_rewards}
\end{figure*}

We additionally estimate the optimal policies for both learning agents using PPO with the default parameters in RLlib. We observe convergence in cumulative training rewards per episode for both methods for the range of values of equitability weight $\eta$ and the GEI weight $\beta$ given in Table \ref{tab:expt_numerics}. Figure \ref{fig:deep_v_tab} is a plot comparing the cumulative training rewards for the learning MM and learning exchange for $\beta\in\lbrace0.0,0.5\rbrace$ and $\eta=1.0$. We see that PPO is able to achieve higher cumulative training rewards for both learning agents. Figure \ref{fig:training_rewards} is a plot of cumulative training rewards achieved using PPO alone for a wide range of values of $(\beta,\eta)$. We see that the cumulative training rewards converge enabling us to estimate optimal exchange fee schedules and MM actions simultaneously for the range of weights $(\eta,\beta)$ considered.

\subsection{Explaining learnt policies}
\begin{figure*}[tb]
    \centering
    \includegraphics[width=\linewidth]{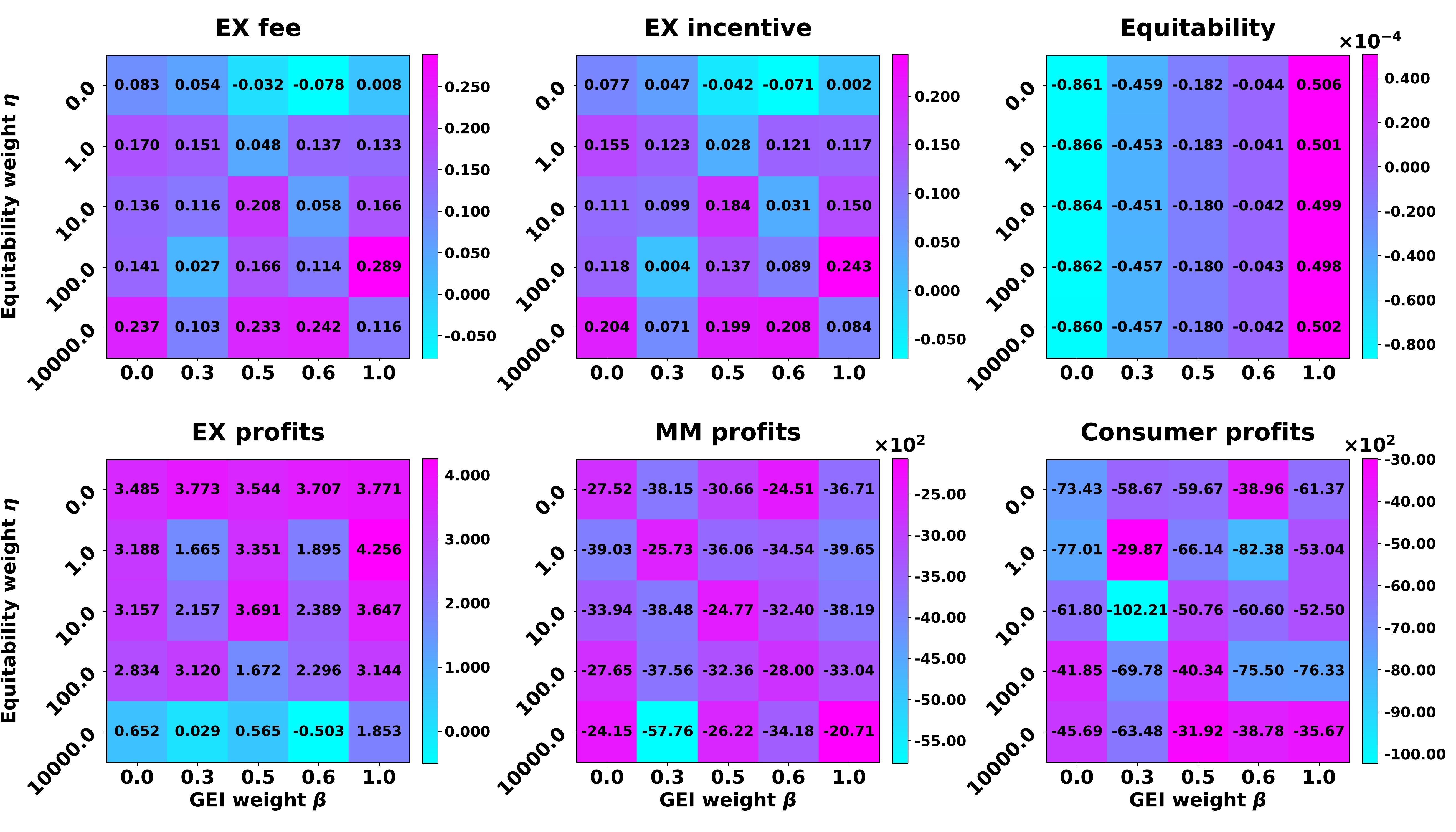}
    \caption{Average policies and profits of MM and exchange with varying $(\eta,\beta)$ for deep Q learning}
    \label{fig:avg_policy}
\end{figure*}



We now try to intuitively explain the effects of the parameters $(\eta,\beta)$ on the learnt exchange and MM policies. Increasing $\beta$ from 0 to 1 corresponds to increasing the weighting of GEI given to MM compared to consumer investors in (\ref{eq:w_gei}), with $\textnormal{GE}_2(Y_{\textnormal{MM}})=0$ with a single MM. While $\beta=0$ accounts for equitability to only the consumer investor group, $\beta=1$ corresponds to the case where the GEI metric captures (between group) equitability to only MMs. On the other hand, increasing $\eta$ corresponds to increasing the equitability weight in the exchange reward (\ref{eq:r_ex}). 

Figure \ref{fig:avg_policy} is a plot of average policies and resulting profits for the EX and MM for various $(\eta,\beta)$ pairs\footnote{All profits are normalized, and hence unit less. EX fee adn incentive are in cents.}. The average policies are got by averaging the learnt policy (which maps the current state to the estimated (optimal) action to be taken in that state) using a uniform distribution on the states. By convention, we are looking at fees charged to liquidity consumers and incentives provided to liquidity providers as in direct stock exchanges. Negative fees correspond to rebates to consumers, and negative incentives correspond to fees charged to providers. Thus, negative fee and negative incentive reflect inverted stock exchanges. We observe the following trends from Figure \ref{fig:avg_policy}. 

\paragraph{Exchange fees and incentives}
As $\eta$ increases for a given $\beta$, we see that the exchange starts charging more fees from liquidity consumers. For some $\beta$, we see that the exchange moves from initially providing incentives to liquidity consumers to charging them. When $\beta$ increases given high values of $\eta$, going from considering equitability to consumer investors to that for MMs, the fees to consumers increase. We see similar trends in the exchange incentives for liquidity providers. As $\eta$ increases for a given $\beta$, we see that the exchange starts providing more incentives to liquidity providers. For some $\beta$, we see that the exchange moves from initially charging fees to liquidity providers to giving them incentives. When $\beta$ increases given high values of $\eta$, going from considering equitability to consumer investors to that for MMs, the incentives to providers increase.

The above two points say that when the exchange is looking at equitability to only consumer investors, increasing the equitability metric makes it switch from an inverted exchange to a direct exchange. This is in line with popular opinion about direct exchanges being more equitable to consumer investors than inverted exchanges. 

\paragraph{Exchange and MM profits}

As $\eta$ increases for fixed $\beta$, we see exchange profits decreasing as it strives to be more equitable. For fixed $\eta$, as $\beta$ is increased to consider equitability to the MM, the exchange profit increases in line with MM profits. Similarly, we see that MM profits increase as $\beta$ is increased to favour equitability to the MM group. 

\paragraph{Consumer profits and equitability}
As the equitability weight $\eta$ increases for a given $\beta$, we see consumer profits increase. For a fixed high value of equitability weight $\eta$, when $\beta$ increases going from considering equitability to consumer investors to that for MMs, we interestingly see that consumer profits increase. This is to say that the MMs are incentivized to provide liquidity to consumers in an equitable fashion. For the equitability reward (\ref{eq:r_ex_equitability}), we see that it increases as the weight to the MM group is increased. This goes to say (as previously) that focusing solely on equitability to the MM group helps in the equitability in the entire marketplace since the MM is then incentivized to provide liquidity in an equitable fashion (at the cost of low exchange profits).

\section{Discussion and Conclusion}
In this paper, we used reinforcement learning to design a dynamic fee schedule for a marketplace agent that makes the marketplace equitable while ensuring profitability for trading agents. 
We see that the choice of equitability parameters define the nature of learnt policies for strategic marketplace agents. The learnt policies start favoring the agent group with the highest equitability weight. We observe that such a setup can be used to design marketplace incentives to wholesale agents to influence them to make marketplaces more equitable. 

\begin{acks}
This paper was prepared for informational purposes by the Artificial Intelligence Research group of JPMorgan Chase \& Co. and its affiliates (``JP Morgan''), and is not a product of the Research Department of JP Morgan. JP Morgan makes no representation and warranty whatsoever and disclaims all liability, for the completeness, accuracy or reliability of the information contained herein. This document is not intended as investment research or investment advice, or a recommendation, offer or solicitation for the purchase or sale of any security, financial instrument, financial product or service, or to be used in any way for evaluating the merits of participating in any transaction, and shall not constitute a solicitation under any jurisdiction or to any person, if such solicitation under such jurisdiction or to such person would be unlawful.
\end{acks}

\printbibliography

\appendix

\end{document}